\documentclass[twoside,fleqn]{article}

\usepackage{graphicx}
\usepackage{espcrc2}
\usepackage{amssymb}
\usepackage{array}

\def\CO{{\ensuremath{\cal O}}}

\def\gsim{{\mathrel{\raise2pt\hbox to 8pt{\raise -5pt\hbox{$\sim$}\hss{$>$}}}}}
\def\rsim{{\mathrel{\raise2pt\hbox to 8pt{\raise -5pt\hbox{$\sim$}\hss{$>$}}}}}
\def\lsim{{\mathrel{\raise2pt\hbox to 8pt{\raise -5pt\hbox{$\sim$}\hss{$<$}}}}}

\def\etal{{\it et al.}}

\title{
       \begin{flushright}\normalsize
	    \vskip -0.9 cm
            LA-UR-01-5861
       \end{flushright}
	\vskip -0.4 cm
	Scaling behavior of improvement and renormalization constants \thanks{
	Presented by Rajan Gupta. Calculations supported by the DoE Grand Challenges award 
	at the ACL at Los Alamos and at NERSC.}
        }
\author{Tanmoy Bhattacharya\address{MS B-285, 
		Los Alamos National Lab, Los Alamos, New Mexico 87545, USA}, 
        Rajan Gupta$\null^{\rm a}$, 
	Weonjong Lee$\null^{\rm a}$,
	Stephen Sharpe\address{Physics Department, University of Washington,
         Seattle, Washington 98195, USA}
       }

\begin{document}

\begin{abstract}
This talk summarizes results for all the scale independent
renormalization constants for bilinear currents ($Z_A$, $Z_V$, and
$Z_S/Z_P$), the improvement constants ($c_A$, $c_V$, and $c_T$), the
quark mass dependence of $Z_{\cal O}$, and the coefficients of the
equation of motion operators for $O(a)$ improved lattice QCD.  Using
data at $\beta=6.0$, $6.2$ and $6.4$ we study the scaling behavior of
these quantities and quantify residual discretization errors.
\end{abstract}

\maketitle


The use of axial and vector Ward identities has proven to be a very
efficient and reliable way of extracting the improvement and
renormalization constants for the $O(a)$ improved fermion action.  The
methodology, references to previous calculations, and the notation we
use are given in ~\cite{LANL:Zfac:00}. The new features of our
calculation summarized here are: new determinations of $c_A$ including
$O(m^2 a^2)$ corrections and using non-zero momentum correlators;
improved chiral extrapolations in the extraction of $Z_A^0$, $c_T$ and
$\tilde b_P\!-\!\tilde b_A$; and a quantitative comparison of the
scaling behavior of the differences between our results and those of
the ALPHA collaboration~\cite{ALPHA:Zfac:97A,ALPHA:Zfac:97A} and
1-loop perturbation theory. Results are summarized in
Table~\ref{tab:finalcomp} and will be presented in detail
in~\cite{LANL:Zfac:01}. \looseness-1

\begin{figure}[t]
\begin{center}
\leavevmode\includegraphics[width=1.0\hsize] {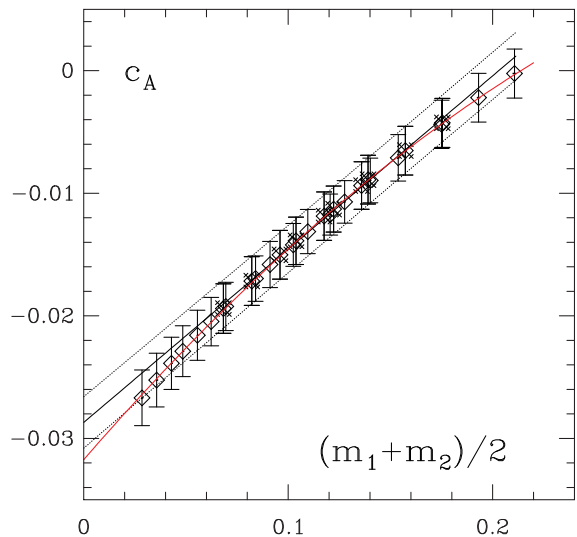}
\vskip -0.9cm
\caption{Comparison of linear and quadratic extrapolation of $c_A$ to the chiral limit}
\vskip -1.5cm
\label{f:cAext64}
\end{center}
\end{figure}

The first feature we discuss is the need for including an $O(m^2 a^2)$ term
in the extrapolation of $c_A$ to the chiral limit. Data at $\beta=6.4$
is shown in Fig.~\ref{f:cAext64}. Table~\ref{tab:finalcomp}
gives results from both the linear and the preferred quadratic fit. 
Our results show a weak dependence of $c_A$ on $\beta$ in the
range $6.0-6.4$, unlike that found by the ALPHA collaboration, but
consistent with the recent results by Collins \etal~\cite{UK:cA:01}. \looseness-1

\begin{table*}[!ht]
\caption{The first error in LANL estimates is statistical, and the
second, where present, corresponds to the difference between using
2-point and 3-point discretization of the derivative in extraction of
$c_A$. Asterisks mark values which include $O(ma)$
corrections in the chiral extrapolations.}
\setlength{\tabcolsep}{1.9pt}
\begin{tabular}{|| >{\scriptsize}c 
	|| >{\scriptsize}l | >{\scriptsize}l | >{\scriptsize}l 
	|| >{\scriptsize}l | >{\scriptsize}l | >{\scriptsize}l 
	|| >{\scriptsize}l | >{\scriptsize}l | >{\scriptsize}l ||}
\hline
\multicolumn{1}{||c||}{}&
\multicolumn{3}{c||}{\(\beta=6.0\)}&
\multicolumn{3}{c||}{\(\beta=6.2\)}&
\multicolumn{3}{c||}{\(\beta=6.4\)}\\
\hline
         & LANL              & ALPHA            & P. Th.
         & LANL              & ALPHA            & P. Th.
         & LANL              & ALPHA            & P. Th.     \\
         &                   &                  &
         &                   &                  &
         &                   &                  &                \\[-12pt]
\hline			     		       
         &                   &                  &
         &                   &                  &
         &                   &                  &                \\[-12pt]
$c_{SW}$ & 1.769             & 1.769            &  1.521
         &  1.614            &  1.614           &  1.481
         &  1.526            &  1.526           &  1.449         \\
         &                   &                  &
         &                   &                  &
         &                   &                  &                \\[-12pt]
\hline			     		       
         &                   &                  &
         &                   &                  &
         &                   &                  &                \\[-12pt]
$Z^0_V$  & $+0.770(1)   $    & $+0.7809(6)$     &  $+0.810$  
         & $+0.7874(4)  $    & $+0.7922(4)(9)$  &  $+0.821$
         & $+0.802(1)  $     & $+0.8032(6)(12)$ &  $+0.830$   \\
$Z^0_A$  & $+0.807(2)(8) $   & $+0.7906(94)$    &  $+0.829$  
         & $+0.818(2)(5)$    & $+0.807(8)(2) $  &  $+0.839$
         & $+0.827(1)(4)$    & $+0.827(8)(1) $  &  $+0.847$   \\
$Z^0_A$* & $+0.802(2)(8) $   &                  &  
         & $+0.815(2)(5)$    &                  &  
         & $+0.822(1)(4)$    &                  &             \\
$Z^0_P/Z^0_S$		     		       				    
         & $+0.842(5)(1)$    &  N.A.            & $+0.956$  
         & $+0.884(3)(1)$    &  N.A.            & $+0.959$
         & $+0.901(2)(5)$    &  N.A.            & $+0.962$    \\
         &                   &                  &
         &                   &                  &
         &                   &                  &                \\[-12pt]
\hline			     		       
         &                   &                  &
         &                   &                  &
         &                   &                  &                \\[-12pt]
$c_A$    & $-0.037(4)(8)$    & $-0.083(5)$      & $-0.013$  
         & $-0.032(3)(6)$    &  $-0.038(4)$     &  $-0.012$
         & $-0.029(2)(4)$    &  $-0.025(2)$     &  $-0.011$   \\
$c_A$*   & $-0.038(4)$       &                  &  
         & $-0.033(3)$       &                  & 
         & $-0.032(3)$       &                  &             \\
$c_V$    & $-0.107(17)(4)$   & $-0.32 (7)$      & $-0.028$  
         & $-0.09 (2)(1)$    &  $-0.21(7)$      &  $-0.026$
         & $-0.08 (1)(2)$    &  $-0.13(5)$      &  $-0.024$   \\
$c_T$    & $+0.063(7)(29)$   &  N.A.            & $+0.020$  
         & $+0.051(7)(17)$   &  N.A.            &  $+0.019$
         & $+0.041(3)(23)$   &  N.A.            &  $+0.018$   \\
$c_T$*   & $+0.076(10)$      &                  &  
         & $+0.059(8)$       &                  & 
         & $+0.051(4)$       &                  &             \\
         &                   &                  &
         &                   &                  &
         &                   &                  &                \\[-12pt]
\hline			     		       
         &                   &                  &
         &                   &                  &
         &                   &                  &                \\[-11pt]
$\tilde b_V$		     		       				    
         & $+1.43(1)(4)$     &  N.A.            & $+1.106$ 
         & $+1.30 (1)(1)$    &  N.A.            &  $+1.099$
         & $+1.24 (1)(1)$    &  N.A.            &  $+1.093$  \\
$b_V$    & $+1.52(1)$        & $+1.54(2)$       & $+1.274$ 
         & $+1.42 (1)$       & $+1.41(2)$       &  $+1.255$
         & $+1.39 (1)$       & $+1.36(3)$       &  $+1.239$  \\
$\tilde b_A-\tilde b_V$	     		       				    
         & $-0.26(3)(4)$     &  N.A.            & $-0.002$  
         & $-0.11 (3)(4)$    &  N.A.            &  $-0.002$
         & $-0.09 (1)(1)$    &  N.A.            &  $-0.002$   \\
$b_A-b_V$	     		       				    
         & $-0.24(3)(4)$     &  N.A.            & $-0.002$  
         & $-0.11 (3)(4)$    &  N.A.            &  $-0.002$
         & $-0.08 (1)(1)$    &  N.A.            &  $-0.002$   \\
$\tilde b_P-\tilde b_S$	     		       				    
         & $-0.06(4)(3)$     &  N.A.            & $-0.066$  
         & $-0.09 (2)(1)$    &  N.A.            &  $-0.062$
         & $-0.090(10)(1)$   &  N.A.            &  $-0.059$   \\
$\tilde b_P-\tilde b_A$	     		       				    
         & $-0.07(4)(5)$     &  N.A.            & $+0.002$  
         & $-0.09(3)(3)$     &  N.A.            &  $+0.001$
         & $-0.12(2)(5)$     &  N.A.            &  $+0.001$   \\
$\tilde b_P-\tilde b_A$*	     		       				    
         & $-0.08(30)$       &                  & 
         & $+0.03(10)$       &                  & 
         & $-0.02(4)$        &                  &             \\
         &                   &                  &
         &                   &                  &
         &                   &                  &                \\[-12pt]
\hline			     		       
         &                   &                  &
         &                   &                  &
         &                   &                  &                \\[-11pt]
$\tilde b_A$		     		       				    
         & $+1.17(4)(8)$     &  N.A.            & $+1.104$ 
         & $+1.19 (3)(5)$    &  N.A.            &  $+1.097$
         & $+1.16 (2)(3)$    &  N.A.            &  $+1.092$   \\
$b_A$		     		       				    
         & $+1.28(3)(4)$     &  N.A.            & $+1.271$ 
         & $+1.32 (3)(4)$    &  N.A.            &  $+1.252$
         & $+1.31 (2)(1)$    &  N.A.            &  $+1.237$   \\
$\tilde b_P$		     		       				    
         & $+1.10(5)(13)$    &  N.A.            & $+1.105$ 
         & $+1.23 (11)(7)$   &  N.A.            &  $+1.099$
         & $+1.13 (4)(7)$    &  N.A.            &  $+1.093$  \\
$\tilde b_S$		     		       				    
         & $+1.16(6)(11)$    &  N.A.            & $+1.172$ 
         & $+1.31 (10)(6)$   &  N.A.            &  $+1.161$
         & $+1.22 (4)(8)$    &  N.A.            &  $+1.151$   \\[3pt]
\hline
\end{tabular}
\label{tab:finalcomp}
\end{table*}

\begin{figure}[t]
\vskip -.05cm
\leavevmode\includegraphics[width=1.0\hsize] {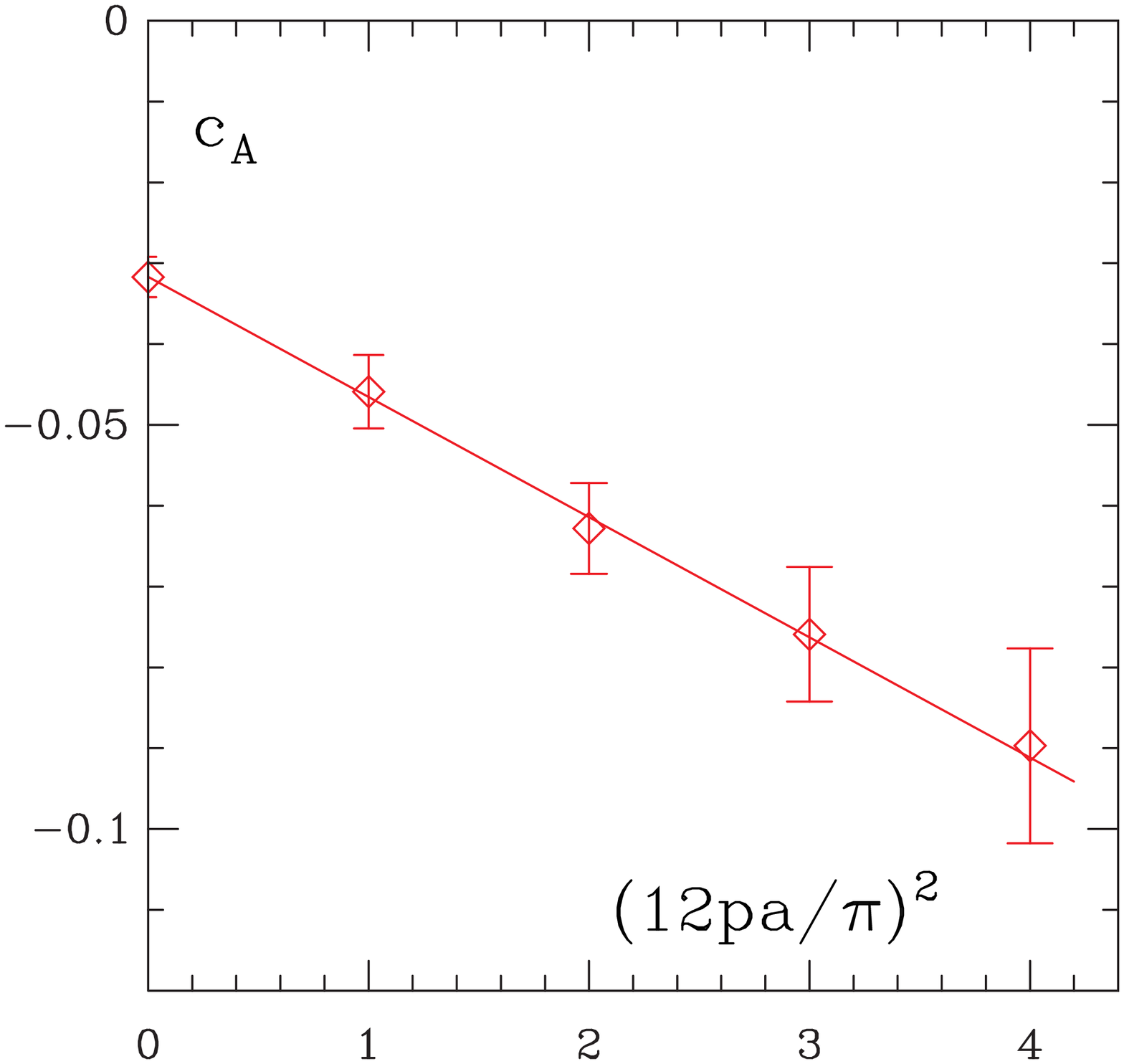}
\vskip -1.0cm
\caption{Evidence for additional $O(p^2a^2)$ errors in $c_A$ extracted 
from $p \neq 0$ correlators.}
\vskip -0.7cm
\label{f:cAmom}
\end{figure}

The second new feature is the demonstration that consistent estimates
for $c_A$ are obtained from correlators with zero and non-zero
momentum once additional $O(p^2a^2)$ errors are accounted for.  A plot
of $c_A$ versus $(12pa/\pi)^2$ at $\beta=6.4$ is shown in
Fig.~\ref{f:cAmom}.  We find that a linear extrapolation to $p=0$
yields results consistent with those obtained using zero momentum
correlators, and with a slope of expected magnitude. \looseness-1

The second point concerns the chiral extrapolation for $Z_A^0$, $c_T$,
$\tilde{b}_P - \tilde{b}_A$.  Our estimates presented
in~\cite{LANL:Zfac:00} were based on constant fits as these quantities
are not expected to have $O(ma)$ corrections if the theory is fully
improved to $O(a)$. We now advocate using results of linear
extrapolation (marked with an asterisk in Table~\ref{tab:finalcomp})
as our data show a dependence on $m$. Such behavior can be explained
by $O(a\Lambda_{QCD} \, ma)$ corrections which can arise as a result
of using a mass-dependent $c_A$ in intermediate stages of
the calculations. To show the size of this effect, we give both estimates
in Table~\ref{tab:finalcomp}.  Even though a fit linear in $ma$
removes only part of the $O(a^2)$ corrections, we choose it as our
preferred value as it is less sensitive to the $m$ values used in the
fit.  One exception is $\tilde{b}_P - \tilde{b}_A$ at $\beta=6.0$ for
which the constant fit is our preferred estimate as the data do not
show a linear term. \looseness-1

Our estimates of $Z_A^0$, $Z_V^0$, $c_A$, $c_V$ differ significantly
from those obtained by the ALPHA collaboration who used the
Schrodinger functional method, while those for $b_V$ agree. We expect
the difference (for example $\Delta Z_V^0 \equiv Z_V^0(LANL) -
Z_V^0(ALPHA)$) to vanish as $O(a^2)$ for $Z_A^0$ and $Z_V^0$, and as
$O(a)$ for $c_A$ and $c_V$. To test this we fit the data assuming
this leading behavior and requiring that the difference vanish at
$a=0$:
\begin{eqnarray}
\label{eq:diffALPHA}
\Delta Z_V^0 &=&   -(55 a)^2\quad - (464 a)^3  \,, \\
\Delta Z_A^0 &=&   -(246 a)^2\ \, + (629 a)^3 \,, \\
\Delta c_A   &=&   -(181 a) + (763 a)^2   \,, \\
\Delta c_V   &=&   -(367 a) + (669 a)^2   \,,
\end{eqnarray}
where $a$ is in units of (MeV${}^{-1}$) and has values $1/2120$,
$1/2910$ and $1/3850$ at the three $\beta$. Considering that the
expected size of the terms is $O(a \Lambda_{QCD})^n$, all the
coefficients look reasonable, however, the errors in them are
large. We make the following observations:
\begin{itemize}
\item
The difference in $Z_V^0$ is dominated by the $O(a^3)$ term. 
\item
The errors in the coefficients for $ Z_A^0 $, and $c_V$ are $> 100\%$.
This is not surprising since the combined error at each $\beta$ is
approximately equal to the difference.
\item
The coefficients in the fit for $\Delta c_A$ have reasonable errors,
however the fit is dominated by the significant difference at
$\beta=6.0$. We note that the $\beta$ dependence of the difference is similar 
for $c_A$, $ Z_A^0 $, and $c_V$. 
\end{itemize}

Using our three data points we can also fit the 
difference between the non-perturbative and tadpole improved 1-loop
estimates as a function of the leading residual discretization error in $a$ 
and perturbative, $O(\alpha_s^2)$, corrections. The results are 
\begin{eqnarray}
\label{eq:diffpert}
\Delta Z_A^0          &=&     -(158 a)^2      -   (1.4 \alpha_s)^2 \\
\Delta Z_V^0          &=&\ \ \, (197 a)^2      -   (1.4 \alpha_s)^2 \\
\Delta Z_P^0/Z_S^0    &=&     -(502 a)^2\,    -   (1.8 \alpha_s)^2 \\
\Delta c_A            &=&     -(13  a)\quad   -   (1.3 \alpha_s)^2 \\
\Delta c_V            &=&     -(51  a)\quad   -   (1.7 \alpha_s)^2 \\
\Delta c_T            &=&\ \ \, (94  a)\quad  +   (0.8 \alpha_s)^2 \\
\Delta {\tilde b_V}   &=&\ \  (930 a)\ \,     -   (2.6 \alpha_s)^2 \\
\Delta b_V            &=&\ \  (429 a)\ \      +   (1.5 \alpha_s)^2 
\end{eqnarray}
where $a$, expressed in MeV${}^{-1}$, $\alpha_s = g^2 /(4\pi u_0^4)$
is the tadpole improved coupling with values $0.1340$, $0.1255$ and
$0.1183$ at the three $\beta$, and $u_0$ is $\langle plaquette
\rangle^{1/4}$. The errors in the other $\tilde b$ are too large to
allow any meaningful fits.

The errors in the coefficients for the three $\Delta Z$'s are
reasonably small, providing some confidence in the fits. Over this
range of $\beta$, the perturbative corrections dominate the
differences in $Z_A^0$ and $Z_V^0$, whereas in $Z_P^0/Z_S^0$, the two
corrections are comparable.

The errors in the coefficients for the three $\Delta c$'s are large.
Even though the coefficients are of the size expected, it is important
to note that the non-perturbative estimates are $2-4$ times the
perturbative values.

Both corrections are large in ${\Delta \tilde b_V}$ and $\Delta b_V$,
with the discretization error being the larger of the two.

Overall, these fits, since they are based on data at just three
$\beta$ values with $1/a$ between $2.1$ and $3.86$ GeV and since we
have ascribed no errors to $a$ or $\alpha_s$, should be considered
indicative and qualitative and certainly not sufficient to draw
precise conclusions. This is why we refrain from quoting errors in the
fits.

\begin{table}
\setlength{\tabcolsep}{3pt}
\renewcommand{\arraystretch}{1.2}
\begin{center}
\caption{Results for off-shell mixing coefficients.}
\begin{tabular}{|c|c|c|c|c|}
\hline
\multicolumn{1}{|c|}{\bf $\beta$}&
\multicolumn{1} {c|}{\bf $6.0$(f)}&
\multicolumn{1} {c|}{\bf $6.0$(b)}&
\multicolumn{1} {c|}{\bf $6.2$} &
\multicolumn{1} {c|}{\bf $6.4$}\\
\hline
 $c'_V+c'_P$  &  $ 2.82 (  15 )  $ &  $ +2.68 (  19 )  $ &  $ 2.62 (   8 )  $ &  $ 2.44 (   4 )  $  \\
 $c'_A+c'_P$  &  $ 2.43 (  24 )  $ &  $ +2.12 (  31 )  $ &  $ 2.43 (  14 )  $ &  $ 2.27 (   6 )  $  \\  
 $2c'_P    $  &  $ 0.88 (  97 )  $ &  $ -0.65 (  57 )  $ &  $ 1.82 (  24 )  $ &  $ 1.85 (   8 )  $  \\   
 $c'_S+c'_P$  &  $ 2.44 (  13 )  $ &  $ +2.40 (  13 )  $ &  $ 2.40 (   7 )  $ &  $ 2.27 (   4 )  $  \\   
 $c'_T+c'_P$  &  $ 2.40 (  18 )  $ &  $ +2.27 (  20 )  $ &  $ 2.42 (   9 )  $ &  $ 2.28 (   5 )  $  \\ \hline
 $c'_V     $  &  $ 2.38 (  50 )  $ &  $ +3.00 (  37 )  $ &  $ 1.72 (  16 )  $ &  $ 1.52 (   4 )  $  \\ 
 $c'_A     $  &  $ 1.99 (  56 )  $ &  $ +2.45 (  46 )  $ &  $ 1.53 (  20 )  $ &  $ 1.35 (   6 )  $  \\ 
 $c'_P     $  &  $ 0.44 (  49 )  $ &  $ -0.33 (  29 )  $ &  $ 0.91 (  12 )  $ &  $ 0.93 (   4 )  $  \\ 
 $c'_S     $  &  $ 2.00 (  48 )  $ &  $ +2.72 (  33 )  $ &  $ 1.49 (  14 )  $ &  $ 1.35 (   4 )  $  \\ 
 $c'_T     $  &  $ 1.96 (  49 )  $ &  $ +2.60 (  38 )  $ &  $ 1.51 (  15 )  $ &  $ 1.36 (   4 )  $  \\   
\hline
\end{tabular}
\vskip -0.9cm
\label{tab:c'}
\end{center}
\end{table}

Finally, in Table~\ref{tab:c'} we present results for the coefficients
of the equation of motion operators. Estimates at $\beta=6.0$ are poor, 
but become reasonably precise at $\beta=6.2$ and $6.4$. We find that
except for $c_P'$, the corrections to the tree level value $c_\CO' = 1$
are large.

%


\end{document}